\documentclass{article}

\usepackage{fancyhdr}
\usepackage{arxiv}

\usepackage[utf8]{inputenc} 
\usepackage[T1]{fontenc}    
\usepackage{hyperref}       
\usepackage{url}            
\usepackage{booktabs}       
\usepackage{amsfonts}       
\usepackage{amsmath}        
\usepackage{nicefrac}       
\usepackage{microtype}      
\usepackage{lipsum}		
\usepackage{float} 
\usepackage{graphicx}
\usepackage{natbib}
\usepackage{doi}

\title{Efficient EEG Seizure Detection Using INT8 Quantization, Channel Pruning, and Spiking Neural Networks}

\author{ {Kartikey Ahlawat}\\
	Leiden Institute of Advanced Computer Science \\
	Leiden University\\
	\texttt{kartikeyprit@gmail.com} \\
}	

\providecommand{\undertitle}{}

\hypersetup{
  pdftitle={Efficient EEG Seizure Detection Using INT8 Quantization, Channel Pruning, and Spiking Neural Networks},
  pdfsubject={EEG seizure detection, model compression, spiking neural networks},
  pdfauthor={Kartikey Ahlawat, Xiangyuan Tao, Jyothis Gireesan Mini, Priyanjali Goel, Hannah Faghihi},
  pdfkeywords={EEG, seizure detection, INT8 quantization, channel pruning, spiking neural networks},
}

\begin{document}
\maketitle

\begin{abstract}
	Continuous EEG monitoring for epilepsy is constrained by the limited power and memory budgets of wearable and implantable devices. Deep neural networks can detect seizures with high accuracy, but their computational cost and model size make them difficult to deploy on such platforms. In this work we use a single 1D CNN seizure detector on the CHB-MIT scalp EEG dataset \cite{chbmit_2010} as a common baseline, and then investigate three brain-inspired efficiency strategies: (i) conversion of the CNN into a spiking neural network (SNN) via parameter transfer, (ii) EEG channel pruning combined with 2{:}4 structured weight sparsity, and (iii) INT8 quantization using FX- and ONNX-based workflows, including quantization-aware training and operator fusion. The quantized CNN variants reduce stored model size from 1.63\,MB to 0.44\,MB, lower estimated energy per inference by up to 64\%, and achieve as much as $2.8\times$ speedup in CPU latency while preserving, and in one case slightly improving, AUC. The pruned CNN halves the number of input channels and non-zero weights with only a modest accuracy drop, and the SNN conversion provides a spiking implementation with sparse temporal activity. Together, these experiments characterize three complementary efficiency directions for seizure detection. The code is available here\footnote{\url{https://github.com/kartikeyahl/brain-inspired-computing}}.
\end{abstract}


\section{Introduction}

Physiological signals such as EEG contain rich information for real-time disease detection, including epileptic seizures. For a wearable EEG monitor to be clinically useful, it must operate continuously for days or weeks on a small battery. Current artificial neural networks (ANNs), although highly accurate, are computationally expensive and power-hungry, making them difficult to deploy for always-on, on-device processing. This power barrier remains a major bottleneck for widespread adoption of next-generation wearable health technology. It motivates brain-inspired approaches that explicitly exploit sparsity, event-driven processing, and low-precision computation hallmark properties of biological neural systems.

\subsection{Brain-Inspired Perspective}

The human brain performs complex computation at roughly 20\,W, far below typical AI systems. Our work is guided by three principles that connect seizure detection to brain-inspired computing.

First, spiking neural networks (SNNs) communicate via sparse, asynchronous spikes rather than dense, synchronous activations~\citep{malcolm2023comprehensive}. This event-driven computation promises substantial energy savings, especially on neuromorphic hardware.

Second, biological networks exhibit both sparse connectivity and synaptic pruning. In a similar spirit, we remove redundant EEG channels and impose structured sparsity on network weights, reducing the number of active inputs and synapses without dramatically degrading performance.

Third, synaptic strengths in the brain are noisy and limited in precision~\citep{hubara2017quantized}. Quantizing network parameters and activations to low-bit representations mimics this constraint and simultaneously reduces memory footprint and arithmetic cost.

\subsection{Goal and Research Question}

The primary research question in this work is:
\emph{How do spiking neural networks (SNNs), network quantization, and channel/weight pruning compare as efficiency strategies for EEG-based seizure detection when each is applied to the same convolutional baseline?}

Our goal is to systematically explore the trade-off between model efficiency (power, memory, latency) and diagnostic performance (AUC, accuracy). By evaluating each technique under a shared data pipeline and architecture, we aim to clarify which directions offer the most promising gains for low-power deployment.

\subsection{Related Work}
Deep neural networks have achieved strong performance in EEG-based seizure prediction and classification, with convolutional models capturing spatial–temporal patterns effectively~\cite{kaziha2020convolutional,abdelhameed2021efficient}, but their computational cost hinders deployment on resource-constrained wearables. To reduce sensing and model complexity, several works focus on channel reduction: permutation-entropy based selection~\cite{ra2021novel}, adaptive CNNs such as SlimSeiz~\cite{Lu2025SlimSeiz}, and explainability-guided pruning with DeepSHAP (SHAP-AAD)~\cite{Salmi2025SHAPAAD} all show that a smaller subset of electrodes can preserve diagnostic accuracy. In parallel, spiking neural networks (SNNs) have been explored for seizure detection as an energy-efficient alternative, including spiking CNNs with discrete spike encoding and cross-patient generalization~\cite{tian2021new,muneeb2023energy} and spiking convolutional transformers~\cite{chen2024epilepsy}. Recent approaches such as SpQuant-SNN and QP-SNN combine low-bit quantization with structured pruning to obtain near-baseline accuracy at substantially lower memory and compute cost~\cite{hasssan2024spquant,wei2025qp}. Low-precision computing more broadly relies on post-training quantization and quantization-aware training to map 32-bit floating-point models to low-bit integer arithmetic~\cite{krishnamoorthi2018quantizing,jacob2018quantization}, often implemented via QDQ-style inference engines that fuse quantize–dequantize operations for throughput gains~\cite{wu2020integer}.

Most of these works focus on a single efficiency axis at a time; in contrast, we compare SNN conversion, channel pruning with structured sparsity, and quantization side by side, all applied to a shared CNN baseline on CHB-MIT.
\vspace{-3mm}
\section{Method / Approach}
The methodology is built upon a single 1D Convolutional Neural Network (CNN) baseline for seizure detection on the CHB-MIT Scalp EEG Dataset. We systematically applied and compared three primary efficiency strategies: (i) conversion to a Spiking Neural Network (SNN) via parameter transfer, (ii) a dual-pruning approach targeting both input channels (18 to 8) and network weights (2:4 structured sparsity), and (iii) hierarchical INT8 quantization, including Static Post-Training Quantization (PTQ), Quantization-Aware Training (QAT), and ONNX runtime optimization, to explore the trade-offs between efficiency and diagnostic performance. All the above use the dataset created in section \ref{sec2.1}

\subsection{Dataset \& Pre-processing}
\label{sec2.1}
We analyse CHB-MIT Scalp EEG Dataset, which is a collection of EEG recordings of 22 pediatric subjects with intractable seizures. The recordings are grouped into 24 cases and were collected from 22 subjects. 

\begin{figure}[H]
    \centering
    \includegraphics[width=0.8\linewidth]{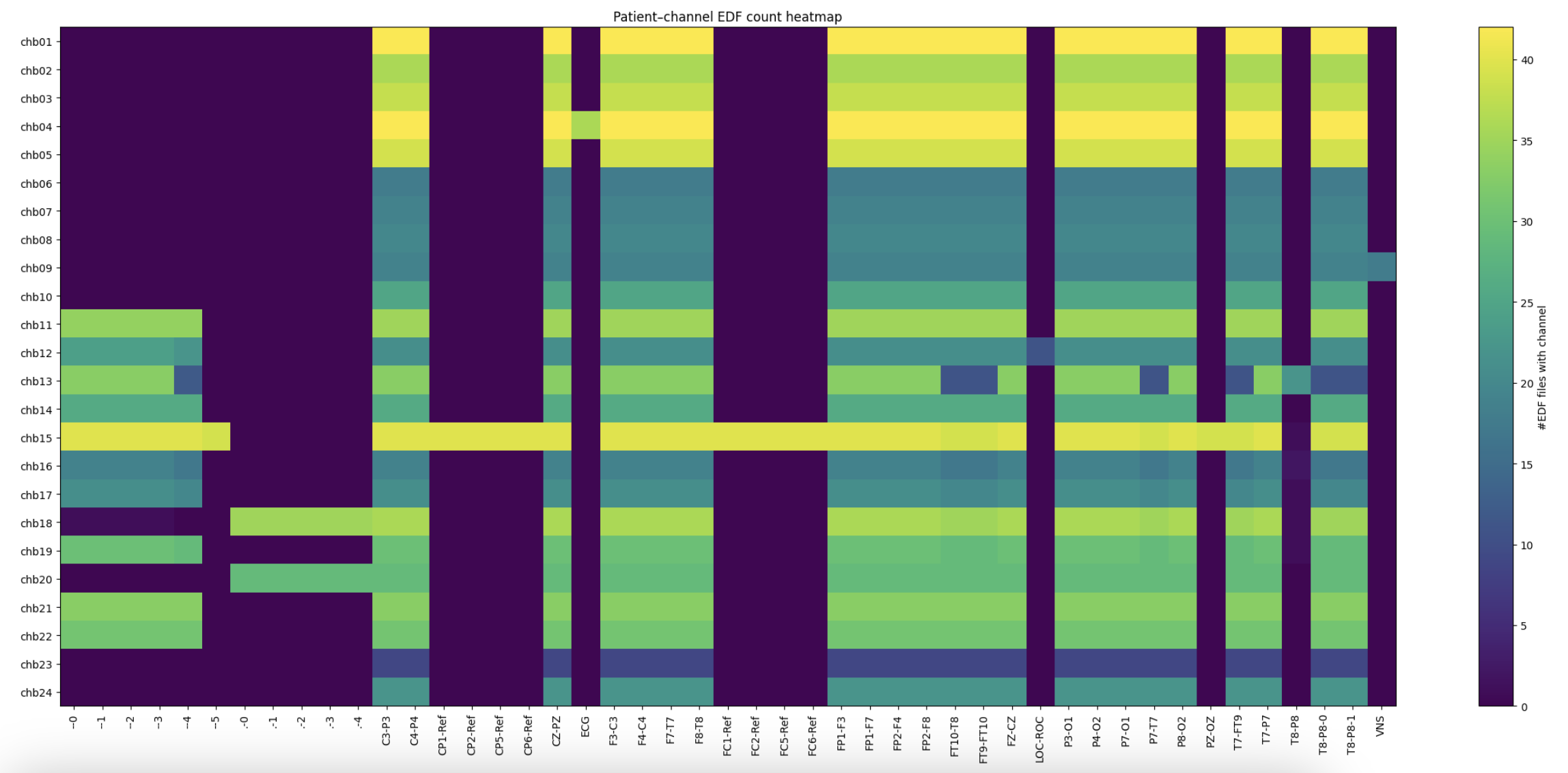}
    \caption{Heatmap on Patient coverage on channels indicating via its number of EDF files}
    \label{fig:channel_distribution}
\end{figure}

\textbf{Channel analysis}: There are a total of 23 channels available in the dataset. However, as seen in the heatmap (Figure \ref{fig:channel_distribution}), there is noticeable sparsity across patients, with some channels missing in several EDF files. To ensure consistent input for model training, we use only the channels that are present in at least 90\% of the patients, which results in selecting 18 channels.
Channel coverage is computed as the ratio between the number of patients containing a given channel and the total number of patients. To obtain a more balanced dataset and avoid missing channel issues during training, we base all experiments on these 18 commonly available channels.
\begin{figure}[H]
    \centering
    \includegraphics[width=0.8\linewidth]{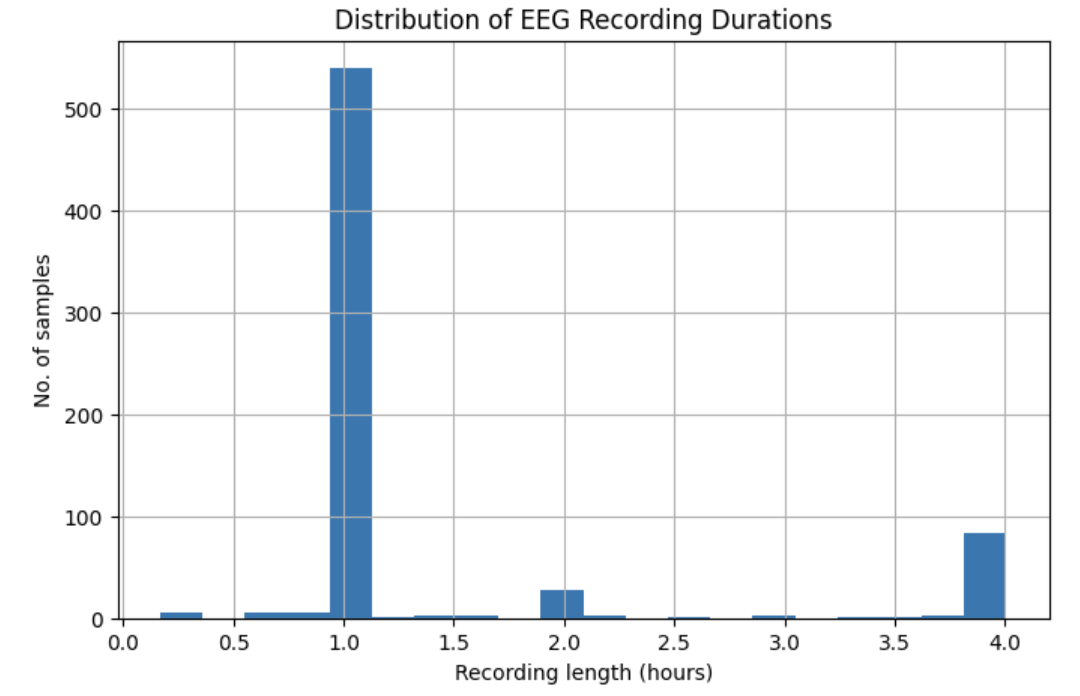}
    \caption{Distribution of EEG Recording with respect to durations.}
    \label{fig:time_data_distribution}
\end{figure}
\textbf{Creation of dataset}: 
Based on the curated list of EEG channels and the available EDF files, we construct a standardized dataset by preprocessing raw recordings, detecting data quality issues, aligning seizure annotations, and generating uniform fixed-length sliding windows suitable for model training. There are a total of about 686 EDF files with 921600 samples (recording per second at frequency of 256Hz)per EDF file.  
\begin{itemize}
\item \textbf{Handling of gaps in EEG files}: The official website of the dataset \cite{chbmit_2010} conveyed the presence of gaps in certain ways for ex. long-term EEG recordings in the CHB-MIT dataset occasionally contain discontinuities. To ensure the temporal consistency of the data, we identified gap samples using two criteria: (i) any sample containing NaNs in any channel, and samples whose across-channel range was flatline). (ii) Samples where the instantaneous amplitude range across channels is below a small threshold (i.e $5\times10^{-6}$). All gap samples were removed before filtering and window extraction.
\item \textbf{Sliding window}: Each recording in the dataset lasts about 1 -4~hours\ref{fig:time_data_distribution}. Since the majority of the recording is for 1 hr, we considered to utilise the segment length  of 8 seconds with a step of 4 seconds to increase the number of training
samples \cite{salmi2025shap}. At a sampling rate of 256 Hz, each window contains 2048 samples (256 Hz × 8 s).
\item \textbf{Seizure annotation}: Seizure annotations are loaded from a corresponding .seizures file . A binary seizure mask of length N is created, in which \textbf{1} indicates that the sample lies inside a seizure interval and 
\textbf{0} indicates a non-seizure sample

\item \textbf{Normalization and Downsampling}: The original 256 Hz recordings were downsampled to 128 Hz to suppress high-frequency drifts\cite{Salmi2025SHAPAAD}. We then applied z-score normalization to standardize the signals and used a band-pass filter to isolate frequency ranges associated with cognitive and affective dynamics \cite{mynoddin2025b}. This resulted in 1024 samples per window (128 Hz x 8 sec).
\end{itemize} 
After preprocessing and windowing, the final dataset consists of 11,145 samples with binary labels indicating a case of seizure/non-seizure. Here, each sample is an 18-channel EEG segment of 1,024 time points. We divide the dataset into training, validation, and test partitions following a standard 60–20–20 split. The training set contains 6,241 samples, the validation set includes 2,675 samples, and the test set comprises 2,229 samples. This split ensures that the model is trained on the majority of the data while retaining sufficient samples for unbiased validation and final performance evaluation.

\subsection{Baseline CNN Architecture}
\begin{figure}[htb]
    \centering
    \includegraphics[width=\textwidth]{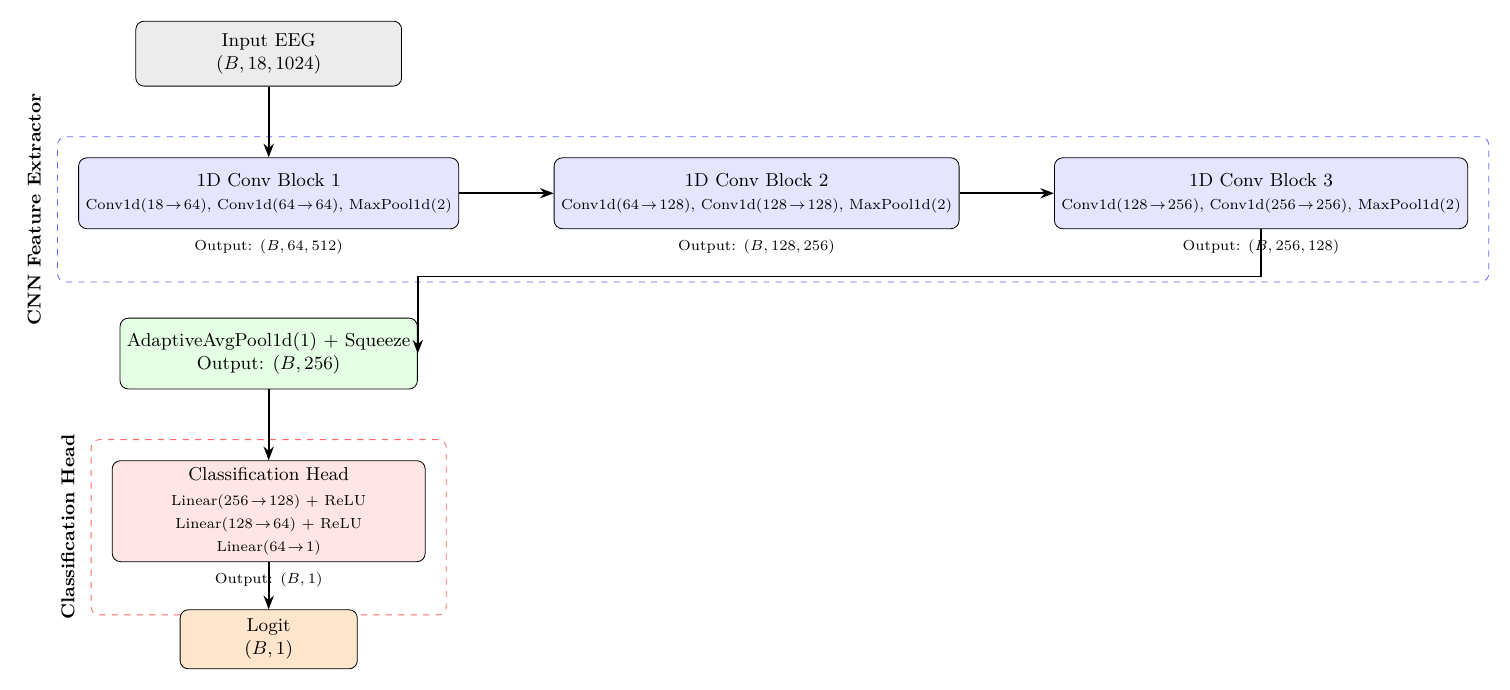} 
    \caption{Baseline 1D CNN Architecture for Seizure Detection.}
\label{fig:cnn_architecture}
\end{figure}
The baseline model is a 1D convolutional neural network operating on the temporal axis. It takes an 18-channel EEG window and passes it through three convolutional blocks of the form Conv1d–ReLU–Conv1d–ReLU–MaxPool1d, with 64, 128, and 256 output channels, respectively (Figure~\ref{fig:cnn_architecture}). An adaptive global average pooling layer then compresses the temporal dimension, and the resulting 256-dimensional vector is fed to a three-layer MLP (256$\rightarrow$128$\rightarrow$64$\rightarrow$1) that outputs a single seizure vs.\ non-seizure logit. The network is trained with binary cross-entropy loss and the Adam optimizer (learning rate $10^{-4}$, batch size 256) for up to 200 epochs with early stopping. This FP32 CNN serves as the reference model for all subsequent SNN, pruning, and quantization experiments.

\subsection{Spiking Network (SNN) via Parameter Transfer}
 
    We convert the 1D CNN into SNN version by transferring its learned weights and biases and replacing ReLU activations with Integrate-and-Fire (IF) neurons \cite{sengupta2019goingdeeperspikingneural}, ensuring the spiking activity approximates the original continuous activations. 
    The main challenge in this conversion is mapping CNN's continuous activation range [0, $Max_{curr}$] to SNN's binary spike (0, 1). For this, we recorded the maximum activation value per layer and then scaled the learned weights (layer-wise weight normalisation) accordingly. This insures the output of every layer remains equal to or lower than the SNN's fixed threshold ($V_{th}=1$ in normalised SNN). The following scaling rule was applied for weight and bias:
    \[
\text{$\text{W}_{new}$} = \text{$\text{W}_{old}$}\times \frac{\text{$\text{Max}_{prev}$}}{\text{$\text{Max}_{curr}$}}
\]
    \[
\text{$\text{b}_{new}$} = \text{$\text{b}_{old}$}\times \frac{1}{\text{$\text{Max}_{curr}$}}
\]
\begin{itemize}
    \item $\text{W}_{new}$ and $\text{W}_{old}$, denote the new weight after scaling and the old weight before scaling, respectively.
    \item $\text{Max}_{prev}$, denotes maximum activation value of the previous layer.
    \item $\text{Max}_{curr}$, denotes the maximum activation value of the current layer.
    \item $b$, signifies the bias term.
\end{itemize}
IF neuron constitutes of the following three custom operations:
\begin{itemize}
    \item Integration: Its purpose is to accumulate the incoming current ($x$) into the membrane potential of the neuron. 
    \[
    mem_{curr}\leftarrow mem + x
    \]
    \item Firing: It is responsible for activating the neuron if the spiking threshold is met.
    \[
    spike = (mem_{curr}\geq V_{th}).float()
    \]
    \item Reset: It ensures to reset the membrane potential after the spike. If the accumulated current in the membrane excceeds the threshold, then it is kept in the membrane for the next iteration. This makes sure that the neurons accumulating higher currents get fired faster.
    \[
    mem = mem_{curr} - spike\times threshold
    \]
\end{itemize}
The SNN was trained for 50 steps with batch size of 32.
    \[
    ConV_1\rightarrow IFNeuron\rightarrow Binary\ Spike\ o/p\rightarrow ConV_2\rightarrow ...
    \]
\subsection{Channel and Weight Pruning}
We implemented a Dual-Pruning Strategy targeting both data redundancy (input level) and model complexity (weight level).

\subsubsection{Structured Channel Pruning (Input Level)}

We employed a Gradient-Sensitivity\cite{Xie2021PruningWithCompensation} Analysis to reduce the dimensionality of the EEG input. Instead of arbitrarily selecting channels, we calculated the sensitivity of the model's loss function with respect to the input features:

\[
S_c = \sum \left| \frac{\partial L}{\partial x_c} \right|
\]

Channels with higher gradient magnitudes \( S_c \) indicate a stronger influence on the decision boundary. We ranked all 18 bipolar channels and retained the Top-8 most discriminative channels, discarding noise and redundant signals.

\subsubsection{Semi-Structured Weight Pruning (Model Level)}

To compress the model parameters, we applied 2:4 Structured Sparsity \cite{Mozaffari2025SLIM}. This technique is hardware-aware (optimized for modern Tensor Cores). For every contiguous block of 4 weights in the convolutional and linear layers, we retained the 2 weights with the largest magnitude and forced the other 2 to zero. This achieves a 50\% reduction in model parameters while maintaining regular memory access patterns.

\subsubsection{Implementation Details}

The pruning pipeline was executed in three phases:

\begin{itemize}
    \item \textbf{Sensitivity Ranking:} We passed a subset of the validation set through the pre-trained baseline model to compute input gradients.

    \item \textbf{Dataset Slicing \& Retraining:} The dataset was sliced to keep only the selected 8 channels. A new CNN model was instantiated with reduced input dimensions and retrained from scratch to adapt to the optimized feature set.

    \item \textbf{Weight Masking:} The 2:4 sparsity mask was computed and applied to the retrained model. We performed a fine-tuning phase to recover any potential accuracy loss caused by weight removal.
\end{itemize}

\subsection{Quantization techniques} 

To bridge the gap between high-performance deep learning and energy-constrained wearable devices, we implemented a hierarchical quantization pipeline. While SNNs offer theoretical efficiency, industry-standard edge hardware (e.g., DSPs, mobile NPUs) is currently optimized for integer arithmetic \citep{jacob2018quantization}. While our original goal included quantizing the Spiking Neural Network (SNN), this proved infeasible within the constraints of current libraries. Popular SNN toolkits such as snnTorch \citep{eshraghian2021snntorch} and Norse \citep{pehle2021norse} do not currently support native integer quantization. Furthermore, extremely low-bit formats require specialized custom kernels \citep{khoram2018adaptive} which are unavailable in common environments. We therefore experimented aggressively on INT8 quantization of the EEG-CNN.

We utilized PyTorch FX Graph Mode \citep{reed2022torchfx} and ONNX Runtime to explore four precision-reduction strategies:

\textbf{Dynamic vs. Static Quantization:} 
We initially applied Dynamic Post-Training Quantization (PTQ), converting weights to INT8 while keeping activations in FP32. However, as dynamic quantization primarily targets Linear layers, compression was limited ($1.07\times$). We transitioned to Static PTQ, which quantizes both weights and activations. We calibrated the model using a representative subset of 512 samples to determine static clipping ranges via min-max observers \citep{krishnamoorthi2018quantizing}.

\textbf{Operator Fusion:} 
To reduce memory access overhead—identified as the primary energy bottleneck in edge inference \citep{horowitz2014computing}—we explicitly fused Convolutional and Linear layers with their subsequent ReLU activations (\texttt{Conv1d+ReLU}). This allows the hardware to perform the activation function immediately in the accumulator without writing intermediate FP32 values back to DRAM.

\textbf{Quantization-Aware Training (QAT):} 
Static quantization can degrade performance on bio-signals due to the high dynamic range of ictal (seizure) spikes, which may be clipped by static observers. To mitigate this, we employed QAT \citep{jacob2018quantization}, simulating quantization noise (fake-quantization modules) during the training phase. This allows the network to learn weights that are robust to the discrete INT8 grid.

\textbf{Inference Engine Optimization:} 
Finally, we exported the models to ONNX (Open Neural Network Exchange) \citep{bai2019onnx} to leverage graph optimizations (constant folding) and the highly optimized CPU execution provider.

To quantify the battery impact without a physical power meter, we employ the "Race-to-Sleep" energy model \citep{horowitz2014computing}. The total energy per inference ($E_{inf}$) is modeled as the product of the processor's active power consumption ($P_{active}$) and the inference latency ($t_{lat}$):

\begin{equation}
    E_{inf} \approx P_{active} \times t_{lat} + E_{mem}
    \label{eq:energy}
\end{equation}

Where $P_{active}$ is the dynamic power of the microcontroller (assumed $\approx 100mW$ for a standard ARM Cortex-M4F DSP). Crucially, \citet{horowitz2014computing} demonstrates that memory access energy ($E_{mem}$) is a key part of computation energy. By reducing the model size from 1.63MB to 0.44MB via quantization, we ensure the model fits entirely within on-chip SRAM, eliminating high-energy off-chip DRAM fetches. Thus, minimizing $t_{lat}$ becomes the primary proxy for minimizing total battery drain.

\section{Results}

\subsection{Baseline with SNN and Quantization}
\begin{table}[h]
\centering
\caption{Quantization \& Energy Results. Energy is estimated using Eq. \ref{eq:energy} assuming a 100mW active power profile. \textbf{Fused Static ONNX} reduces energy consumption by 64\% compared to the baseline, while \textbf{Fused QAT} achieves the highest accuracy.}
\vspace{0.5em} 
\label{tab:quant_results}
\resizebox{\textwidth}{!}{%
\begin{tabular}{l|c|c|c|c|c|c}
\hline
\textbf{Model Config} & \textbf{Precision} & \textbf{Size (MB)} & \textbf{Latency (ms)} & \textbf{Est. Energy ($\mu$J)} & \textbf{Speedup} & \textbf{Test AUC} \\ \hline
Baseline CNN & FP32 & 1.63 & 0.39 & 39.3 & 1.0x & 0.9617 \\
SNN & FP32 & 1.63 & 112.5 & 11,225 & -288x & 0.9258 \\
Dynamic PTQ & INT8 (Weights) & 1.52 & 0.54 & 54.1 & 0.7x & 0.9617 \\
Static FX PTQ & INT8 & \textbf{0.44} & 0.26 & 26.1 & 1.5x & 0.9609 \\
\textbf{Fused QAT FX} & \textbf{INT8} & \textbf{0.44} & 0.25 & 25.1 & 1.6x & \textbf{0.9628} \\
Static ONNX (Standard) & INT8 & \textbf{0.44} & 0.21 & 21.0 & 1.9x & 0.9617 \\
\textbf{Fused Static ONNX} & \textbf{INT8} & \textbf{0.44} & \textbf{0.14} & \textbf{13.9} & \textbf{2.8x} & 0.9615 \\ \hline
\end{tabular}%
}

\end{table}
\begin{itemize}
    \item \textbf{Energy Efficiency:} 
As shown in Table \ref{tab:quant_results}, the \textbf{Static ONNX} configuration is the clear winner for battery optimization. By reducing latency to \textbf{0.14ms}, it lowers the estimated energy cost to just \textbf{13.9 $\mu$J} per inference, a \textbf{63\% reduction} compared to the FP32 baseline (39.3 $\mu$J). This implies that for a battery-operated wearable device, the ONNX model allows the processor to return to sleep mode nearly $3\times$ faster than the baseline.
\item  \textbf{Accuracy Analysis:}
Standard Static PTQ caused a minor AUC drop (0.9617 $\to$ 0.9609). However, \textbf{Fused QAT} outperformed the baseline (AUC \textbf{0.9628}). This confirms that simulated quantization noise acts as a regularizer \citep{nagel2021white}, improving robustness to noisy EEG signals.\\
SNN became the least performing model out of all. Achieved $AUC: 0.92$ after 50 time steps. Unlike CNN, it requires multiple iterations to accumulate current and create a spike. These multiple iterations increase the inference latency and if steps are reduced then AUC decreases (trade-off). SNN's true performance advantage is unlocked only on specialized neuromorphic hardware. This hardware executes the spike updates asynchronously and when required instead of sequential n-step loop on a CPU, reducing latency and energy.

\end{itemize}
\subsection{Baseline with Pruning}
The proposed dual-pruning strategy resulted in substantial reductions in computational overhead: the input bandwidth decreased by 55\% (from 18 to 8 channels), and the model storage requirements were reduced by 50\% through 2{:}4 structured weight sparsity.

\begin{table}[H]
    \centering
    \caption{Performance and efficiency comparison between the Baseline (18-channel) and the final Pruned (8-channel, 2:4 sparse) model.}
        \vspace{0.5em} 
    \label{tab:pruning_results}
    \begin{tabular}{l|c|c|c}
        \hline
        \textbf{Metric} & \textbf{Baseline CNN} & \textbf{Pruned} & \textbf{Change} \\ \hline 
        Accuracy & 96.17\% & 95.15\% & -1.02\% \\ 
        Input Channels & 18 & 8 & -55\% \\ 
        Model Weights & 100\% (Dense) & 50\% (Sparse) & -50\% \\ \hline
    \end{tabular}
    
\end{table}

Despite these aggressive reductions, the impact on classification performance was minimal. The pruned model achieved an accuracy of 95.15\%, corresponding to a negligible drop of 1.02\% relative to the baseline.

\section{Conclusion}

We evaluated three efficiency strategies i.e SNN conversion, pruning, and INT8 quantization—on a 1D CNN for EEG seizure detection. INT8 Quantization specifically optimizes the digital processing stage. By transitioning to the Fused Static ONNX engine, we reduced inference latency by $2.8\times$ (0.39ms $\to$ 0.14ms), enabling an aggressive ``Race-to-Sleep'' strategy. This cut the estimated processor energy per classification by 64\% (from 39.3$\mu$J to 13.9$\mu$J), drastically extending the device's battery life.

SNN conversion was the least efficient on standard CPUs, showing a 288$\times$ latency increase and the largest AUC drop (0.9258), confirming that SNN benefits require neuromorphic hardware.

Pruning on baseline reduced both input and model complexity i.e channel pruning decreased dimensionality from 18 to 8 channels, and 2:4 structured sparsity removed 50\% of weights, with minor drop in accuracy drop, effective in the cases of lowering computational load.  

\section{Future Work}
Quantization methods can be improved further by bridging QAT robustness with ONNX efficiency via the \textbf{QDQ export pipeline} \citep{wu2020integer}, theoretically merging peak accuracy (0.9628 AUC) with minimal energy (13.9$\mu$J). Additionally, exploring \textbf{INT4 quantization} to further halve the memory footprint, utilizing mixed-precision strategies to preserve bio-signal fidelity in sensitive layers.

Beyond quantization alone, an important direction is to \emph{combine} the three efficiency axes studied here. One concrete target is an 8-channel CNN that is simultaneously 2{:}4 sparse and INT8-quantized, providing a single compact model for edge CPUs. In parallel, applying structured pruning and weight sharing directly to the SNN variant, followed by deployment on neuromorphic hardware, would allow a more realistic comparison in terms of synaptic operations and true power consumption. Finally, patient-specific fine-tuning and extension to multi-class seizure-type classification would move these brain-inspired models closer to practical clinical use.

\bibliographystyle{unsrtnat}
\bibliography{references}  

\end{document}